\documentclass[12pt,preprint]{aastex}         
\def\be{\begin{equation}}
\def\ee{\end{equation}}

\def\bb{\mbox{\boldmath $\beta $}}

\def\lsim{\lower 2pt \hbox{$\, \buildrel {\scriptstyle <}\over
         {\scriptstyle \sim}\,$}}

\begin{document}
\newcommand{\figureout}[2]{ \figcaption[#1]{#2} }       

\title{Effects of Rotation and Relativistic Charge Flow on Pulsar Magnetospheric Structure}

\author{Alex G. Muslimov\altaffilmark{1,2} \& 
Alice K. Harding\altaffilmark{2}}   

\altaffiltext{1}{Present address: ManTech International Corporation, 
Lexington Park, MD 20653}

\altaffiltext{2}{Laboratory of High Energy Astrophysics,      
NASA/Goddard Space Flight Center, Greenbelt, MD 20771}
 

\begin{abstract}
We propose an analytical 3-D model of the open field-line region of a neutron star (NS) magnetosphere. 
We construct an explicit analytic solution for arbitrary obliquity (angle between the rotation 
and magnetic axes) incorporating the effects of magnetospheric rotation, relativistic flow of 
charges (e.g. primary electron beam) along the open field lines, and ${\bf E}\times {\bf B}$-drift 
of these charges. Our solution employs the space-charge-limited longitudinal current calculated 
in the electrodynamic model of Muslimov \& Tsygan (1992) and is valid up to very high altitudes 
nearly approaching the light cylinder. We assume that in the innermost magnetosphere, the 
NS magnetic field can be well represented by a static magnetic dipole configuration. 
At high altitudes the open magnetic field lines significantly deviate from those of a static 
dipole and tend to focus into a cylindrical bundle, swept back in the direction opposite to the 
rotation, and also bent towards the rotational equator.  We briefly discuss some implications of our 
study to spin-powered pulsars.     
\end{abstract} 


\pagebreak
  
\section{INTRODUCTION}

The growing observational data (from radio- to $\gamma -$ rays) on spectra and pulse profiles 
of spin-powered pulsars prompt continued improvement of theoretical models (see e.g. Harding 2005, Kaspi et al. 
2004). For example, study of particle acceleration and radiation produced at high altitudes in a pulsar 
magnetosphere (Muslimov \& Harding 2004a [MH04], Hirotani et al. 2003) 
depends heavily on our knowledge of the structure of open magnetic flux lines (passing through 
the light cylinder) at very high altitudes, where the standard static magnetic dipole approximation is no longer accurate. Thus, the calculation reported here was undertaken initially with the specific purpose of a quantitative description of the distorted magnetic field in a realistic magnetosphere at high altitudes 
. In this paper we develop the corresponding analytic solution which can be 
used up to very high altitudes nearly approaching the light cylinder. 

Since the pioneering theoretical studies (Goldreich \& Julian 1969; Ostriker \& Gunn 1969; Sturrock 1971; 
Mestel 1971; Ruderman \& Sutherland 1975; and Arons \& Scharlemann 1979) of pulsar phenomena there continues 
to be interest in the magnetospheric structure of a rotating neutron star (NS). The equation governing the 
structure of an axisymmetric pulsar magnetosphere was derived (under quite strong assumptions and idealizations) 
more than three decades ago (see e.g. Mestel 1973; Scharlemann \& Wagoner 1973; Michel 1973; Okamoto 1974; Mestel et al. 1979 and references therein) and can be reduced to the well-known (special-relativistic) force-free 
Grad-Shafranov equation (see Grad 1967; Shafranov 1966 for generic version of the equation). Since then the basic 
ideas of these and similar studies have shaped the school of thought that seeks to construct a mathematically closed (albeit highly idealized and axisymmetric) model of a NS magnetosphere and wind zone. The contemporary development of this school is mostly represented by Mestel and collaborators (see Mestel 1999 for general overview; and also Goodwin et al. 2004 for the most recent version) and by the Lebedev Institute group (Beskin et al. 1983). Recently, Bogovalov (2001) studied the MHD plasma flow in the magnetosphere of an oblique rotator with an initially split-monopole magnetic field. However, his solution cannot directly apply to radio pulsars since it is 
valid when $R_a < R_{\rm lc}$ (where $R_a$ and $R_{\rm lc}$ are the Alfven and light cylinder radius, respectively). Thus, despite significant progress in the numerical solution (see Contopoulos et al. 1999; Mestel 1999) of the Grad-Shafranov pulsar equation, numerical simulation of plasma in a rotating NS magnetosphere (Biltzinger \& Thielheim 2004; and Spitkovsky 2004), and also in the computation of MHD winds (Bogovalov 2001 and references therein), it is difficult to find in the literature any useful estimate of the high-altitude distortion of open-field lines of e.g. initially dipolar magnetic structure. How does the magnetospheric  distortion at high altitude depend on the pulsar obliquity? Is the space-charge-limited current sufficient to distort open-field lines at high altitudes, and if so, how will it change the form of the open-field-line bundle? What is the high-altitude radial dependence of B? None of the existing NS magnetosphere models can readily provide clear and simple answers to these and similar questions. The main reason is that these models do not have simple analytic versions. 
The only available analytic model is the classical ``vacuum" model of Deutsch (1955) which is not applicable (see also Section 3.1) to the physical situation in a real pulsar magnetosphere filled with charges and currents.   

In this paper we approach the problem in a slightly different way by first identifying and understanding the main physical effects distorting the geometry of the open-field line configuration within the light cylinder. Then, taking  advantage of the fact that these effects enter Maxwell's equations either as the first or second order terms in the radial distance, scaled by the light cylinder radius, we can significantly facilitate the problem by separating the multiple terms in the coupled system of equations. In doing so, we are able to solve analytically the simplified Maxwell's equations to determine the corrections to the static magnetic field caused by each of these effects. The main element of our model is our use of the electric current along the open field lines, which is determined by the 
space-charge-limited flow solution in the electrodynamic model of Muslimov \& Tsygan (1992 [MT92]). We try to keep our formalism as simple as possible, so that our solution can be easily reproduced. The solution presented in this paper, being mathematically more transparent, allows us to understand the general picture of how the open magnetic field lines of a NS get distorted at high altitudes, depending on the pulsar obliquity, on the distribution of the electron current over the polar cap (PC) and therefore on the magnetic latitude and azimuth, and on the ${\bf E} \times {\bf B}$-drift of charges. More importantly, the presented solution illustrates the way each of the effects of rotation, charge flow and ${\bf E}\times {\bf B}$ -drift contributes to the resulting pattern of distorted open field lines. Thus, the treatment is an attempt to explore a more realistic situation that is intermediate to the extreme cases of vacuum and MHD studied in the past. Although in the present study we focus on the region of pulsar magnetosphere confined by the light cylinder, we understand that the particles streaming along the open field lines will form a relativistic wind zone (Mestel et al. 1979). The regime of relativistic wind and corresponding global configuration of the magnetic field will be discussed separately.

The paper is organized as follows. In \S ~2 we present a set of fundamental electrodynamic equations that will 
be employed in our study. In \S ~3 we formulate our approach and discuss how to incorporate the effect of 
rotation and relativistic charge flow (\S ~3.1), and the effect of ${\bf E}\times {\bf B}$-drift of electrons 
(positrons) on the structure of open field line region within the light cylinder of a NS 
magnetosphere (\S ~3.2). In \S ~4 we provide 3-D views of open field lines for different obliquities. 
Finally, in \S ~ 5 we discuss the results of our study and summarize its most exciting implications for 
spin-powered pulsars.

\section{Basic Equations}

Let us consider the magnetosphere of a rotating NS and assume that in the frame of reference rigidly 
corotating with the NS the magnetic field is stationary. The very general equations describing the 
electromagnetic field produced by the rotating NS in the Lab (inertial) frame are the first couple,
\be
{\bf \nabla} \cdot {\bf B} = 0,
\label{divB}
\ee
\be
{\bf \nabla} \times {\bf E} = - {1\over c} {{\partial {\bf B}}\over {\partial t}},
\label{curlE}
\ee
and the second couple of Maxwell's equations,
\be
{\bf \nabla } \cdot {\bf E} = 4 \pi \rho ,
\label{divE}
\ee
\be
{\bf \nabla}\times {\bf B} = {1\over c} {{\partial {\bf E}}\over {\partial t}} + 
{{4 \pi}\over c} {\bf j},
\label{curlB}
\ee
where $\rho $ and $\bf j$ are the electric charge and current density, respectively.
 
In this paper we adopt the standard picture that the NS's magnetosphere has two distinctive regions: 
a `dead zone' with field lines that close within the light cylinder, and without any current flow 
along the field; and the open-field line region extending beyond the maximum radius of corotation 
(in cylindrical coordinates with the $z$-axis along the NS's rotation axis), the light-cylinder radius $R_{\rm lc}\approx c/\Omega $, and with flow of charges along the field lines. As in our previous 
studies, we will be working in a spherical polar coordinate system, ($\eta \equiv r/R, \theta , \phi $), in which 
the polar axis is parallel to the magnetic moment.  In this system, $r$ is the radial coordinate and $R$ is the 
stellar radius, $\theta$ is the polar angle measured from the magnetic dipole axis, and $\phi$ is the azimuthal 
angle measured counter-clockwise from the meridian passing through the rotation axis.  We refer to this coordinate 
system throughout the paper as magnetic coordinates.  Finally, we define $\chi$ to be the pulsar obliquity (angle 
between the NS rotation axis and magnetic dipole moment).

In our model calculation we will assume that the static (unperturbed by rotation and currents) 
magnetic configuration of the NS has a pure dipole geometry,
\be
{\bf B}^{\rm d} = {{B_0^{\rm d}}\over \eta ^3}~\left(~\cos \theta ~{\bf e}_{\rm r} + {1\over 2}~
\sin \theta ~{\bf e}_{\theta }~\right),
\label{Bd}
\ee
where ${\bf e}_{\rm r}$, ${\bf e}_{\theta }$ are the corresponding basis vectors of the magnetic
coordinate system defined above and $B_0^{\rm d}$ is the magnetic field strength at the magnetic pole. 
Here, for the sake 
of simplicity, we ignore static general relativistic corrections to the magnetic field. This is well justified, 
because we are interested in the corrections to the magnetic field at high altitudes approaching the light cylinder 
and caused by the magnetosphere rotation and flow of charges along the open field lines. The only exception 
will be the expression for the longitudinal component of the current density (see e.g. formula [\ref{j||}] below) 
which is essentially determined by the condition at the stellar surface where general relativistic effects are 
not merely important but make a qualitative difference.

\section{Formulation of the problem and analytic solution}

\subsection{The effect of rotation and charge flow}

We will be searching for the steady-state solution to equations (\ref{divB})-(\ref{curlB}). In this 
case, the time derivatives in equations (\ref{curlE}) and (\ref{curlB}) are determined by the rotation of 
the NS magnetosphere relative to the Lab frame. It is important that well within the light cylinder 
($\eta \lsim \eta _{\rm lc} \equiv R_{\rm lc}/R$) this rotation is most likely a solid-body rotation, 
and we can use the following transformations of partial time derivatives between the Lab frame (subscript ``Lab") 
and the frame of reference corotating with NS magnetosphere (subscript ``corot"):
\be
\left\{ {{\partial {\bf B}}\over {\partial t}}  \right\} _{\rm corot} = \left\{ {{\partial {\bf B}}\over {\partial t}}  \right\} _{\rm Lab} - {\bf \nabla } \times ({\bf u}_{\rm rot} \times {\bf B}),
\label{dB/dt}
\ee
\be
\left\{ {{\partial {\bf E}}\over {\partial t}}  \right\} _{\rm corot} = \left\{ {{\partial {\bf E}}\over {\partial t}}  \right\} _{\rm Lab} - {\bf \nabla } \times ({\bf u}_{\rm rot} \times {\bf E}) + {\bf u}_{\rm rot}~{\bf \nabla \cdot E},
\label{dE/dt}
\ee
\be
\left\{ {{\partial \rho }\over {\partial t}}  \right\} _{\rm corot} = \left\{ {{\partial \rho }\over {\partial t}}  \right\} _{\rm Lab} + {\bf u}_{\rm rot} \cdot {\bf \nabla} \rho,
\label{drho/dt}
\ee
where ${\bf u}_{\rm rot}$ is the rotational velocity of the magnetosphere.

We assume that in a steady state, the time derivatives in the LHS of equations (\ref{dB/dt})-(\ref{drho/dt}) 
vanish, so that Maxwell's equations (\ref{curlE}) and (\ref{curlB}) can be rewritten in the following form
\be
\nabla \times {\bf E} = - \nabla \times ( {\bb }_{\rm rot} \times {\bf B}),
\label{curlE-2}
\ee
\be
\nabla \times {\bf B} = \nabla \times ( {\bb }_{\rm rot} \times {\bf E}) - 
{\bb}_{\rm rot} \nabla \cdot {\bf E} + {{4\pi}\over c}{\bf j}. 
\label{curlB-2}
\ee
where ${\bb}_{\rm rot} = {\bf u}_{\rm rot}/c$.

\noindent
Also, the charge continuity equation,
\be 
{{\partial \rho }\over {\partial t}} + \nabla \cdot {\bf j} = 0,
\label{drho/dt-2}
\ee
with the help of relationship (\ref{drho/dt}) takes the form
\be
{\bf u}_{\rm rot }\cdot \nabla \rho - \nabla \cdot {\bf j} = 0.
\label{drho/dt-3}
\ee

By combining equations (\ref{divE}) and (\ref{curlB-2}), we get
\be
\nabla \times \left( {\bf B} - {{{\bf u}_{\rm rot}}\over c} \times {\bf E} \right) = 
{{4\pi}\over c} ( {\bf j} - \rho {\bf u}_{\rm rot}). 
\label{curlB-3}
\ee
Thus, the steady-state solution for a rotating magnetosphere is determined by equations (\ref{divB}), 
(\ref{curlE-2}), (\ref{drho/dt-3}) and (\ref{curlB-3}). Note that, since 
$\nabla \cdot {\bf u}_{\rm rot} = 0$ and $\nabla \cdot (\rho {\bf u}_{\rm rot}) = {\bf u}_{\rm rot} 
\cdot \nabla \rho $, equation (\ref{drho/dt-3}) translates into
\be
\nabla \cdot ( {\bf j} - \rho {\bf u}_{\rm rot}) = 0.
\label{divj}
\ee   
Now we should discuss the physical origin of the current density $\bf j$. Within the light cylinder the 
current is mostly determined by the longitudinal (owing to the relativistic electrons 
streaming along the magnetic field lines) and rotational (owing to the bulk rotational motion of charges) 
components,
\be
{\bf j} = {\bf j}_{\parallel } + {\bf j}_{\rm rot}.
\label{jtot}
\ee
In this Section we ignore the effect of ${\bf E}\times {\bf B}$-drift. This effect will be discussed 
separately in Section 3.2. The main reason is that the net current produced by the ${\bf E}\times {\bf B}$-drift 
may or may not vanish depending on the specific scenario of particle acceleration within the open field line 
region. For example, if the relativistic beam is quasineutral, then charges of both sign will be drifting in 
the same direction and with the same velocity thus producing zero net current. On the contrary, if the beam 
is charged (e.g. primary electrons and quasi-neutral electron-positron plasma are flowing in the region 
with open field lines), then the ${\bf E}\times {\bf B}$-drift of electrons (positrons) can significantly 
contribute to the net current and therefore affect the structure of the magnetic field at high altitudes 
(see Section 2.2 for details). 

\noindent 
The longitudinal and rotational components of the current density can be written as
\be
{\bf j}_{\parallel } = j_{\parallel } {{\bf B} \over B}
\label{jpar}
\ee
and
\be
{\bf j}_{\rm rot} = - |\rho | {\bf u}_{\rm rot},
\label{jrot}
\ee
respectively.

Note that in equation (\ref{jrot}) the ``-" sign signifies that negative charges (electrons) are involved in 
rotational motion.
        
By using expressions (\ref{jpar}) and (\ref{jrot}) we can rewrite equation (\ref{curlB-3}) as
\be
\nabla \times \left( {\bf B} - {\bb}_{\rm rot} \times {\bf E} \right) = 
{{4\pi }\over c} {\bf j}_{\parallel }.
\label{curlB-4}
\ee 
Although this equation is essentially the same as the corresponding equation derived by Beskin et al. (1983), we 
should point out that there are some principal differences: Beskin et al. neglected the component 
${\bf j}_{\rm rot}$ in their expression for $\bf j$; and, in their derivation of the equation (see their eq. [15]) 
analogous to our eq. (\ref{curlB-3}), they neglected the component $\rho {\bf u}_{\rm rot}$ resulting from the 
transformation (\ref{dE/dt}).

We assume that at higher altitudes where the effect of rotation becomes increasingly important, the electric field 
(in the Lab frame) is mostly determined by rotation (see also MH04),
\be
{\bf E} \approx - {\bb}_{\rm rot} \times {\bf B}^{\rm d}.
\label{E}
\ee 
Note that, besides ${\bf j}\neq 0$ and $\rho \neq 0$, relationship (\ref{E}) assures the fundamental difference 
between ours and Deutsch's solutions. Also, we should point out that for high altitudes, the classical vacuum 
analytic solution of Deutsch transforms into a pure wave-like solution well beyond the light cylinder and 
is hardly applicable to the realistic situation. Finally, the Deutsch's solution is presented in 
spherical coordinates with the polar axis along the rotation axis (see also Cheng et al. 2000), whereas 
our solution will be presented in magnetic coordinates. 

\noindent 
Then, equation (\ref{curlB-4}) can be rewritten as
\be
\nabla \times {\bf B} = {{4 \pi }\over c} {\bf j}_{\parallel } + {\bf J},
\label{curlB-5}
\ee
where
\be
{\bf J} = \nabla \times {\bf G},
\label{J}
\ee
and
\be
{\bf G} = {\bb}_{\rm rot}\times {\bf E} \equiv \beta _{\rm rot}^2~{\bf B}^{\rm d}- {\bb}_{\rm rot} 
~({\bb}_{\rm rot } \cdot {\bf B}^{\rm d}).
\label{G}
\ee
Here we assumed that $\bf E$ is determined by equation (\ref{E}).

The main goal of the present study is to construct the appropriate analytic solution for the open magnetic field 
lines valid within the light cylinder. By inspecting the Maxwell's equations derived above one can see that the corresponding solution for vector $\bf B$ can be generally presented as
\be
{\bf B} = {\bf B}^{\rm d} + {\bf B}^{(1)} + {\bf B}^{(2)} + {\bf B}^{(2\ast )},
\label{B}
\ee 
where ${\bf B}^{\rm d}$ is a pure dipole component (see formula [\ref{Bd}]), ${\bf B}^{(1)}$ is the first-order 
correction to the static dipole component which is $\sim (\eta /\eta _{\rm lc})~B^{\rm d}$, and ${\bf B}^{(2)}$ and 
${\bf B}^{(2\ast)}$ are the second-order corrections, $\sim (\eta /\eta _{\rm lc})^2~B^{\rm d}$, respectively. The 
correction ${\bf B}^{(1)}$ is produced by the charge flow along the open field lines. ${\bf B}^{(2)}$ is the 
distortion caused by rotation of ${\bf B}^{\rm d}$, and ${\bf B}^{(2\ast )}$ is the distortion generated by 
${\bf E}\times {\bf B}$ drift of the outflowing charges. The terms ${\bf B}^{\rm d}$ and ${\bf B}^{(2)}$ 
represent a rotating vacuum solution subject to the condition of equation (\ref{E}). 

In what follows, for the sake of convenience, we will use the magnetic spherical coordinates with the polar 
axis along the magnetic dipole moment. 

Obviously, the first-order correction to the dipole magnetic field is generated by the longitudinal current 
flowing along the poloidal (and mostly determined by the dipolar component) magnetic field. In this case the 
contribution from the displacement current $\bf J$ is of the second order and can be neglected. Thus, the 
equation for determining ${\bf B}^{(1)}$ reduces to
\be
\nabla \times {\bf B}^{(1)} = {{4\pi }\over c} {\bf j}_{\parallel },
\label{B1}
\ee
To complete the formulation, we should add the following couple of equations (see equations [\ref{divB}] 
and [\ref{divj}])
\be
\nabla \cdot {\bf B}^{(1)} = 0,
\label{divB1}
\ee
and
\be
\nabla \cdot {\bf j}_{\parallel } = 0.
\label{divj||}
\ee
To solve the system of equations (\ref{B1}) - (\ref{divj||}), we need the explicit expression for 
${\bf j}_{\parallel }$. For this purpose we will employ the electrodynamic model of MT92 and write
\be
{\bf j}_{\parallel } = - c |\rho | {{{\bf B}^{\rm d}}\over {B^{\rm d}}} \approx 
- {\Omega \over {2\pi }}~\left[ (1-\kappa )~\cos \chi + 
{3\over 2}~\theta _0~\xi ~\sin \chi ~\cos \phi \right]~{\bf B}^{\rm d}, 
\label{j||}
\ee
where $\theta _0 \approx (\Omega R/c)^{1/2}$ is the canonical PC half-angle, $\xi $ is the 
dimensionless magnetic colatitude of open field lines ($\xi = 1$ corresponds to the last open field 
lines, and $\xi = 0$ corresponds to the magnetic axis), and $\kappa $ is the parameter 
measuring the general-relativistic effect of frame dragging at the stellar surface in units of stellar angular 
velocity $\Omega $. According to our estimate, for most more or less realistic NS equations of state, 
$\kappa \approx 0.15~I_{45}/R_6^3$; where $I_{45}=I/10^{45}$ g$\cdot $cm$^2$, $R_6=R/10^6$ cm, $I$ is the 
moment of inertia of NS of radius $R$. 

One can easily verify that expression (\ref{j||}) for ${\bf j}_{\parallel }$ satisfies equation 
(\ref{divj||}). Note also that in formula (\ref{j||}) the explicit $\phi $-dependence implies that 
for pure dipole field lines the azimuthal coordinate $\phi $ coincides with the azimuthal coordinate 
of a stream line. In other words, for any given value of $\phi $ one can calculate the current density 
${\bf j}_{\parallel }$ which is fixed at the stellar surface at the azimuth $\phi _{\rm pc}^0~(\equiv \phi )$.   

Let us now introduce the dimensionless vector, ${\bf b}^{(1)}$, such that 
\be
{\bf B}^{(1)} = {\left( {\Omega R}\over c\right) ~B_0^{\rm d}}~{\bf b}^{(1)} (\eta, \theta, \phi ),
\label{b1}
\ee
where $B_0^{\rm d}$ is the dipole field strength at the magnetic pole. 
Then, equation (\ref{B1}) reduces to
\be
{{\partial }\over {\partial \theta }} \left( b_{\phi }^{(1)}~\sin \theta \right) 
- {\partial \over {\partial \phi }} b_{\theta }^{(1)} = - {2\over {\eta ^2}} (\alpha + \beta \cos \phi ) 
\sin \theta \cos \theta ,
\label{curl-r}
\ee
\be
{1\over {\sin \theta }} {{\partial b_{\rm r}^{(1)}}\over {\partial \phi }} - 
{{\partial }\over {\partial \eta }} \left( \eta ~b_{\phi }^{(1)}\right) = - {1\over {\eta ^2}} 
(\alpha + \beta \cos \phi ) \sin \theta ,
\label{curl-theta}
\ee
\be
{{\partial }\over {\partial \eta }}\left( \eta ~b_{\theta }^{(1)} \right) - 
{{\partial b_{\rm r}^{(1)}}\over {\partial \theta }} = 0.
\label{curl-phi}
\ee
where $\alpha \equiv (1-\kappa) \cos \chi $, and $\beta \equiv (3/2) \theta _0 \xi \sin \chi $. 
 
By examining the system of equations (\ref{curl-r}) - (\ref{curl-phi}) and (\ref{divB1}) one can 
see that the solution for ${\bf b}^{(1)}$, that vanishes at infinity, should have the following simple form
\be
b_r^{(1)} = {{f(\theta )}\over {\eta ^2}} \sin \phi , 
\label{br-1}
\ee
\be
b_{\theta }^{(1)} = {{g(\theta )}\over {\eta ^2}} \sin \phi ,
\label{btheta-1}
\ee
\be
b_{\phi }^{(1)} = {1\over {\eta ^2}} [ h_1 (\theta ) + h_2 (\theta ) \cos \phi ].
\label{bphi-1}
\ee
This solution should be regular at the magnetic pole (at $\theta = 0$) and satisfy the periodic condition,  
${\bf b}^{(1)}(\phi ) = {\bf b}^{(1)}(\phi +2\pi)$. By substituting expressions (\ref{br-1})-(\ref{bphi-1}) 
into equations (\ref{divB1}), (\ref{curl-r})-(\ref{curl-phi}), we find that equation (\ref{curl-r}) is a 
consequence of equations (\ref{divB1}), (\ref{curl-theta}) and (\ref{curl-phi}). Also, the solution for 
$h_1(\theta )$ can be written immediately, 
\be
h_1(\theta ) = - \alpha \sin \theta ,
\label{h1}   
\ee
whereas solution for $f$, $g$, and $h_2$ can be found from the system
\be
f'' \sin ^2 \theta + f' \sin \theta \cos \theta - f = \beta \sin ^2 \theta ,
\label{f}
\ee
\be
g = - f',
\label{g}
\ee
\be
h_2 = - {f \over \sin \theta } - \beta \sin \theta ,
\label{h2}
\ee
where $' \equiv \partial /\partial \theta $.
The solution of equation (\ref{f}) that is finite at $\theta \rightarrow 0$ reads
\be
f = \beta (1 - \theta \cot \theta ),
\label{f-sol}
\ee
so that $g$ and $h_2$ can be easily determined after inserting this solution into equations (\ref{g}) and (\ref{h2}), 
respectively. 

Thus, the final analytic solution for ${\bf B}^{(1)}$ can be written as
\be
B_r^{(1)} = {3\over 2} \left( { \eta \over {\eta _{\rm lc}}} \right)~{{B_0^{\rm d}}\over {\eta ^3}}~\theta _0 ~\xi ~ 
( 1 - \theta \cot \theta ) \sin \chi \sin \phi ,
\label{B1r}
\ee
\be
B_{\theta }^{(1)} = - {3\over 2} \left( { \eta \over {\eta _{\rm lc}}} \right)~{{B_0^{\rm d}}\over {\eta ^3}}~
\theta _0~\xi ~\left( { {\theta - \sin \theta \cos \theta }\over {\sin ^2 \theta } }\right) \sin \chi \sin \phi ,
\label{B1theta}
\ee
\be
B_{\phi }^{(1)} = - \left( { \eta \over {\eta _{\rm lc}}} \right)~{{B_0^{\rm d}}\over {\eta ^3}}~
\left[ ( 1 - \kappa ) \cos \chi \sin \theta + {3\over 2} \theta _0 \xi 
\left( {{\sin ^3 \theta + \sin \theta - \theta~\cos \theta } \over {\sin ^2 \theta }} \right) 
\sin \chi \cos \phi \right] .
\label{B1phi}
\ee

Now by substituting into equation (\ref{curlB-5}) the component ${\bf j}_{\parallel }^{(2)}$ determined by 
${\bf B}^{(1)}$ and using the explicit expression for $\bf J$, we arrive at an equation for determining the 
component ${\bf B}^{(2)}$,
\be
\nabla \times {\bf B}^{(2)} = {{4 \pi }\over c}~{\bf j}_{\parallel }^{(2)} + {\bf J},
\label{B2} 
\ee
where 
\be
{\bf j}_{\parallel }^{(2)} = - {\Omega \over {2\pi }}~\Lambda _0~{\bf B}^{(1)}, 
\label{j||2}
\ee
and 
\be
\Lambda _0 = (1-\kappa )~\cos \chi + (3/2)~\theta _0~\xi ~\sin \chi ~\cos \phi _{\rm pc}^0.
\label{Lambda}
\ee
Note that $\Lambda _0$ (cf. expression [\ref{j||}]) depends on $\phi _{\rm pc}^0$ not $\phi $, simply 
because formula (\ref{j||2}) implies that the stream lines of current determining ${\bf B}^{(2)}$ 
now have an azimuthal component, in which case the coordinate $\phi \neq \phi _{\rm pc}^0$ and 
${\bf j}_{\parallel }^{(2)}$ can only be set by specifying its value (or the value of 
$\Lambda _0$) at magnetic azimuth $\phi _{\rm pc}^0$ at the PC surface. Here, for the sake 
of simplicity, we assume that in formula (\ref{Lambda}) the parameters are such that $\Lambda _0 > 0$, 
i.e. the charges of the same sign (electrons) can be ejected from the stellar surface. The case of 
$\Lambda _0 < 0$ is briefly discussed in the last paragraph of Section 4. 
  
Obviously, we can search for a solution for the component ${\bf B}^{(2)}$ as having the following dependence 
of the dimensionless vector ${\bf b}^{(2)}$, 
\be
{\bf B}^{(2)} = \left( { {\Omega R}\over c } \right) ^2~B_0^{\rm d}~{\bf b}^{(2)}.
\label{b2} 
\ee
Note also, that the vector $\bf J$ can be presented as
\be
{\bf J} = \left( {{\Omega R}\over c} \right) ^2~B_0^{\rm d}~ \nabla \times {\bf g}, 
\label{J-2}
\ee 
where the spherical components of vector $\bf g$ are given in Appendix A.  

By using expressions (\ref{j||2}), (\ref{J-2}) and employing the solution for ${\bf B}^{(1)}$, 
equation (\ref{B2}) can be reduced to
\be
\nabla \times {\bf b}^{(2)} = \nabla \times ({\bf g}+{\bf h}),
\label{curl-b2}
\ee
where the spherical components of vector ${\bf h}$ are also given in Appendix A.

From equation (\ref{curl-b2}) we can get
\be
b_r^{(2)} = g_r + h_r. 
\label{b2r} 
\ee
The $\theta -$ and $\phi -$ components of ${\bf b}^{(2)}$ can be presented as
\be
b_{\theta }^{(2)} = g_{\theta } + {1\over \eta } {{\partial X} \over {\partial \theta }} + 
{{C_1(\theta )}\over \eta },
\label{b2-theta}
\ee
\be
b_{\phi }^{(2)} = g_{\phi } + h_{\phi } + {1\over {\eta \sin \theta }} {{\partial X}\over {\partial \phi }} 
+ {{C_2(\phi )} \over {\eta \sin \theta }}.
\label{b2-phi}
\ee
Thus, ${\bf b}^{(2)}$ vanishes at large radial distances, and $X(\theta , \phi )$, $C_1(\theta )$, and $C_2(\phi )$ are some functions to be determined from equation 
\be
\nabla \cdot {\bf b}^{(2)} = 0,
\label{divb2}
\ee
which translates into
\begin{eqnarray}
\eta (g_r+h_r) &+& {1\over {\sin \theta }} \left\{ {{\partial }\over{\partial \theta }} \left[ 
\sin \theta \left( \eta g_{\theta } + C_1 + {{\partial X}\over {\partial \theta }} \right) \right] + \right. 
\nonumber \\
& & \left. {1\over {\sin \theta }} \left[ \eta {\partial \over \partial \phi } 
\left( g_{\phi } + h_{\phi } \right) + {1\over {\sin \theta }} 
\left( {{\partial ^2 X}\over {\partial \phi ^2}} + 
{{\partial C_2}\over {\partial \phi }} \right) \right] \right\} = 0. 
\label{divb2-2}
\end{eqnarray}
By separating the variables in equation (\ref{divb2-2}) and assuming the regularity of solution at the 
magnetic pole and periodicity over azimuthal coordinate (see also the comment following equation [\ref{bphi-1}]), 
we can find that
\be
C_1 = - \left\{ {3\over 8}~\left[ \sin ^2\chi ~(1+\cos^2\theta)+2~\cos ^2\chi ~\sin^2\theta \right] +
\Lambda _0~(1-\kappa)~\cos\chi \right\}~\sin \theta ,
\label{C1}
\ee
\be
C_2 = 0,
\label{C2}
\ee
and
\be
X = (U~\sin \chi~\cos \chi + V~\Lambda _0~\theta _0~\xi ~\sin \chi )~\cos \phi + 
W~\sin ^2 \chi~ \cos 2 \phi,
\label{X}
\ee
where
\be
U = {1\over {\sin \theta }}~(\cos \theta - 1) - {1\over 2}~\sin \theta ~\left(\cos ^2 \theta - {1\over 2}\right),
\label{U}
\ee
\be
V = {3\over 2}\left( \cos \theta + {{3~\theta }\over {\sin \theta }} - 4  \right) ,
\label{V}
\ee
and
\be
W = {1\over 8}~\left[ {5\over 2}~\left( {{1 - \cos \theta }\over {\sin ^2 \theta }} - 
{1\over 2} \right) - \sin ^2 \theta ~\cos \theta \right].
\label{W}
\ee
By using the above expressions, we can now write the components of ${\bf B}^{(2)}$,
\begin{eqnarray}
B_{\rm r}^{(2)} & = & \left( {\eta \over {\eta _{\rm lc}} }\right )^2~{{B_0^{\rm d}}\over {\eta ^3}}~
\cos \theta ~\left\{ \cos \chi ~[\cos \chi ~\sin ^2 \theta + 2~\Lambda _0~(1-\kappa )] + 
\sin ^2\chi~\left( 1 - {1 \over 2}~\sin ^2 \theta \right) - \right. \nonumber \\
& & \left. 2~\left[ \cos \chi ~\sin \theta~ \cos \theta + {3\over 2}~\Lambda _0~\theta _0~\xi ~
\left( {\theta \over {\sin \theta ~\cos \theta } } - 1 \right) \right] ~\sin \chi ~\cos \phi - \right. \nonumber \\
& & \left. {1\over 2} ~ \sin ^2 \chi ~\sin ^ 2 \theta~\cos 2\phi \right\},
\label{B2r}
\end{eqnarray}
\begin{eqnarray}
B_{\theta }^{(2)} & = & - \left( {\eta \over {\eta _{\rm lc}}} \right) ^2~{{B_0^{\rm d}}\over {\eta ^3}}~
\sin \theta \left\{ \left[ {1\over 4}\cos \chi ~\sin ^2 \theta + \Lambda _0~(1-\kappa ) \right]~\cos \chi + \right. \nonumber \\
& & \left. {1\over 2}~\sin ^2 \chi~\left( 1 - {1\over 4}~\sin ^2 \theta \right) + \left[ \left( 
{1\over 4}~\cot \theta~\cos 2\theta + {{1-\cos \theta }\over {\sin ^3 \theta }} \right)~\cos \chi - 
\right. \right. \nonumber \\
& & \left. \left. {3\over 2}~\Lambda _0~\theta _0~\xi ~\left( 
{{ 3(1-\theta ~\cot \theta )}\over {\sin ^2\theta }} -1 \right) \right]~\sin \chi~\cos \phi - 
\right. \nonumber \\
& & \left. {5\over 8} \left[ {1\over {\sin ^2 \theta }} \left( 
{{1-\cos \theta }\over {\sin ^2 \theta }} - {1\over 2} \right) + {1\over 5} \sin ^2 \theta \right]~
\sin ^2 \chi~\cos 2\phi~\right\}, 
\label{B2theta}
\end{eqnarray}
\begin{eqnarray}
B_{\phi }^{(2)} & = & \left( {\eta \over {\eta _{\rm lc}}} \right) ^2~{{B_0^{\rm d}}\over {\eta ^3}}~
\left\{ \left[ \left( {1\over 4} + {{1-\cos \theta }\over {\sin ^2 \theta }} \right) ~\cos \chi + 
{3\over 2}~ \Lambda _0~\theta _0~\xi~{{3~\sin \theta ~\cos \theta - \theta ~(3-2~\sin ^2\theta )}\over {\sin ^2 \theta }} \right] \right. \times \nonumber \\
& & \left. \sin \chi ~\sin \phi - {5\over 8}~\left( {{1-\cos \theta }\over {\sin ^3 \theta}} - 
{1\over {2~\sin \theta }}\right) ~\sin ^2\chi ~\sin 2\phi \right\},
\label{B2phi}
\end{eqnarray}

\subsection{Effect of ${\bf E}\times {\bf B}$ drift}

The current density associated with the electron drift within the region of open field lines 
can be written as
\be
{\bf j}_{_{{\bf E}\times {\bf B}}} = - |\rho _0| {c\over {B^2}} {{\bf E}\times {\bf B}},
\label{jEXB}
\ee
where
\be 
|\rho _0| = {\Omega \over {2\pi c}}~\Lambda _0 ~ {{B_0^{\rm d}} \over {\eta ^3}},
\label{rho0}
\ee
from equation (\ref{j||}).
By using the approximation (see [\ref{E}]) ${\bf E}\approx - {\bb}_{\rm rot} \times {\bf B}$, we can 
rewrite expression (\ref{jEXB}) as,
\be
{\bf j}_{_{{\bf E}\times {\bf B}}} = - |\rho _0|~c~{\bb}_{\rm rot} + 
|\rho _0|~c~{\bb}_{\rm rot, \parallel },
\label{jEXB-2}
\ee
where ${\bb}_{\rm rot, \parallel } = ({\bf B}^{\rm d}/B^2)~({\bb }_{\rm rot} 
\cdot {\bf B}^{\rm d} )$, and $B \approx B^{\rm d}$

The second term in (\ref{jEXB-2}) adds to the component of current 
density ${\bf j}_{\parallel } \propto {\bf B}^{\rm d}$ (see eq. [\ref{j||}]). Here we have a situation 
where non-relativistic (or even nearly relativistic) drift motion in the longitudinal direction (along the 
magnetic field lines) is superposed on essentially relativistic flow. Thus, without any loss of generality, we 
can justifiably assume that this component of drift motion can be ignored in the longitudinal 
relativistic flow of charges. However, the first term in ${\bf j}_{_{{\bf E}\times {\bf B}}}$ should be 
explicitly added to ${\bf j}_{\parallel }^{(2)}=-|\rho _0|~c~{\bf B}^{(1)}/B$ (see equation [\ref{B2}]), so 
that the corresponding correction to the magnetic field, ${\bf B}^{(2\ast )}$, produced by the 
${\bf E}\times {\bf B}$-drift current will be determined by the equation
\be
\nabla \times {\bf B}^{(2\ast )} = {{4\pi } \over c}~{\bf j}^{(2\ast )},
\label{B2*}
\ee      
where
\be
{\bf j}^{(2\ast )} = - |\rho _0|~c~{\bb}_{\rm rot }
\label{j2*}
\ee
is the current density associated with the ${\bf E}\times {\bf B}$ drift of electrons.

Now we can introduce the dimensionless vector ${\bf b}^{(2\ast )}$ via the formula
\be
{\bf B}^{(2\ast )} = \left( {{\Omega R}\over {c}} \right) ^2~{B_0^{\rm d}}~\Lambda _0~{\bf b}^{(2\ast )}.
\label{b2*} 
\ee
Then equation (\ref{B2*}) translates into
\be
R~\nabla \times {\bf b}^{(2\ast )} = {\bf i}^{(2\ast )},
\label{curlb2*}
\ee
where vector ${\bf i}^{(2\ast )}$ has the components
\be
i_{\rm r}^{(2\ast )} = 0,
\label{i*r}
\ee
\be
i_{\theta }^{(2\ast )} = {2\over {\eta ^2}}~\sin \chi ~\sin \phi ,
\label{i*theta} 
\ee
and
\be
i_{\phi }^{(2\ast )} = - {2\over {\eta ^2}}~(\cos \chi ~\sin \theta - \sin \chi ~\cos \theta ~\cos \phi ).
\label{i*phi}
\ee
The analytic solution for equation (\ref{curlb2*}) satisfying $\nabla \cdot {\bf b}^{(2\ast )} = 0$ can be easily 
written as
\be
b_{\rm r}^{(2\ast )} = - {2\over \eta }~(\cos \chi ~\cos \theta + \sin \chi ~\sin \theta ~\cos \phi ),
\label{b2*r}
\ee
\be
b_{\theta }^{(2\ast )} = {1\over \eta }~(\cos \chi ~\sin \theta - \sin \chi ~\cos \theta ~\cos \phi )
\label{b2*theta}
\ee
and
\be
b_{\phi }^{(2\ast )} = {1\over \eta }~\sin \chi \sin \phi .
\label{b2*phi}
\ee
Thus, the components of vector ${\bf B}^{(2\ast )}$ can be explicitly presented as
\be
B_{\rm r}^{(2\ast )} = - 2~\left( {\eta \over {\eta _{\rm lc} }} \right) ^2~{{B_0^{\rm d}}\over {\eta ^3}}~\Lambda _0~
(\cos \chi ~\cos \theta + \sin \chi ~\sin \theta ~\cos \phi ),
\label{B2*r}
\ee
\be
B_{\theta }^{(2\ast )} = \left( {\eta \over {\eta _{\rm lc} }} \right) ^2~{{B_0^{\rm d}}\over {\eta ^3}}~
\Lambda _0~(\cos \chi ~\sin \theta - \sin \chi ~\cos \theta ~\cos \phi )
\label{B2*theta} 
\ee
and
\be
B_{\phi }^{(2\ast )} = \left( {\eta \over {\eta _{\rm lc} }} \right) ^2~{{B_0^{\rm d}}\over {\eta ^3}}~\Lambda _0~
\sin \chi ~\sin \phi .
\label{B2*phi}
\ee
Now, by inserting the corresponding components determined by formulae (\ref{Bd}), (\ref{B1r}) - (\ref{B1phi}), 
(\ref{B2r}) - (\ref{B2phi}), and (\ref{B2*r}) - (\ref{B2*phi}) into the general expression (\ref{B}), we can 
calculate the structure of the open magnetic field lines all the way from the NS surface up to the high altitudes 
nearly approaching the light cylinder radius.

\section{Visualization of Analytical Formulae}
  
By using the spherical components of $\bf B$ given by the general expression (\ref{B}), 
we can easily produce a 3-D plot 
of any open field line for arbitrary obliquity. In Figures 1-3 we depict some views of a set of open field lines 
emanating from the same magnetic colatitude of the NS PC and equally spaced (by $30^{\circ }$) in magnetic azimuth.  
We use the Cartesian coordinates X, Y and Z with the center at the magnetic pole. Also, the Z-axis is along the magnetic moment, and the positive X-axis is pointing towards the rotation axis. Finally, all the axes are scaled by the light cylinder radius, and a pulsar spin period of 0.3 s is used. In Fig. 1 we show the side views of open field lines having dimensionless magnetic colatitude $\xi = 0.9$ for different obliquities $\chi = 0$, 30, 60, and $90^{\circ }$. [Here, for the sake of simplicity, we normalize the magnetic colatitude by the canonical PC half-angle, $\theta _0$. In general, the foot points of last open-field lines should be determined by tracing back from the light cylinder to the stellar surface along the last open-field lines. For the purpose of the present study this exercise is not essential, though.] In Fig. 2 we present the same set of magnetic field lines as in Fig. 1 but viewed from the magnetic pole. For the aligned case ($\chi = 0$), as one can see from Figures 1 and 2,  the field lines are axisymmetric and get swept in the direction opposite to rotation, in contrast to the vacuum case. The non-vanishing $B_{\phi }$ component (see equation [\ref{Bphi-aligned}] below) gives a sweep back due solely to 
the real current, even in the aligned case. 
For the aligned rotator the expressions (\ref{B1r}) - (\ref{B1phi}), (\ref{B2r}) - (\ref{B2phi}), and (\ref{B2*r}) - (\ref{B2*phi}) significantly simplify, so that for the components of $\bf B$ defined by formula (\ref{B}) we get
\be
B_{\rm r} = {{B_0^{\rm d}}\over {\eta ^3}}~\cos \theta \{ 1 + (\eta /\eta _{\rm lc})^2~[\sin ^2\theta - 
2~\kappa~(1-\kappa )]\},
\label{Br-aligned}
\ee
\be
B_{\theta } = {1\over 2}~{{B_0^{\rm d}}\over {\eta ^3}}~\sin \theta \{ 1 - (1/2)~(\eta /\eta _{\rm lc})^2~
[\sin ^2\theta - 4~\kappa~(1-\kappa )]\},
\label{Btheta-aligned}
\ee
\be
B_{\phi } = - {{B_0^{\rm d}}\over {\eta ^3}}~(1-\kappa )~(\eta /\eta _{\rm lc})~\sin \theta .
\label{Bphi-aligned}
\ee
From formulae (\ref{Br-aligned}), (\ref{Btheta-aligned}) one can see that at $\theta < \theta _{\rm b}$ (where 
$\theta _{\rm b} \approx [2\kappa (1-\kappa)]^{1/2} \approx 30^{\circ }$) the open field lines have slightly more 
flaring than those of a pure dipole, whereas at $\theta > \theta _{\rm b}$ they are more focused towards
the magnetic axis. Also, in a small-angle approximation ($\theta \ll 1$), from equations (\ref{Br-aligned}), 
(\ref{Bphi-aligned}) we can write the following approximate field-line formula, $\phi \approx \phi _{\rm pc}^0 
- (1-\kappa)~(\eta /\eta _{\rm lc})$. In Figure 3 we illustrate the effect of sweep-back for the aligned 
rotator and for the field lines emanating from different magnetic colatitudes ($\xi = 0.2$, 0.4, 0.5, and 0.9). 
This Figure shows that the effect of sweep-back vanishes towards the magnetic axis. Note that if terms depending 
on the current (last terms in equations [\ref{Br-aligned}] and [\ref{Btheta-aligned}] and the whole RHS of 
equation [\ref{Bphi-aligned}]) are turned off, there still remains a contribution from displacement current. This 
contrasts to the Deutsch solution, where the displacement current $\propto \partial {\bf E}/\partial t \propto 
\partial {\bf E}/\partial \phi $ is zero in the aligned case. This is because the wavelike solution imposed 
by Deutsch at large distances requires non-axisymmetry in order to produce a not vanishing displacement current.           
	
As the obliquity increases (see cases $\chi = 30$ and $60^{\circ }$ in Figures 1 and 2) the leading (negative Y) 
and trailing (positive Y) field lines become asymmetric (with respect to rotation by 180 $^{\circ }$ around 
magnetic axis). Also, in 
this case (Fig. 1, $\chi = 30$ and $60^{\circ }$) and at high altitudes, the field lines get more focused and 
the entire bundle bends away from rotation axis. For the orthogonal rotator (case $\chi = 90^{\circ }$ in 
Figures 1 and 2) and for the value of spin period (0.3 s) we used in our numerical calculation, both the effect of relativistic flow and NS rotation significantly diminish, so that the configuration of open field lines (at least 
up to $\sim 0.3~\eta _{\rm lc}$) is practically the same as in the case of a static magnetic dipole. This is 
difficult to illustrate for arbitrary values of $\theta $ for open field lines, because for $\chi = 90^{\circ }$ 
the analytic formulae still look rather cumbersome. However, in a small-angle approximation, for $\chi = 90^{\circ }$, it is easy to see that $B_{\rm r} \sim B_{\rm r}^{\rm d}$, $B_{\theta } \sim B_{\theta }^{\rm d}$, and 
$B_{\phi } \sim \theta _0~(\eta /\eta _{\rm lc})^2~B_{\rm r}^{\rm d}$, which explains the patterns depicted in 
Figures 1 and 2 for the case of $\chi = 90^{\circ }$. This behavior sharply contrasts to the case of the vacuum 
orthogonal rotator, where the field-line sweep-back is a maximum (Arendt \& Eilek 1998; Cheng et al. 2000; and 
Dyks \& Harding 2004).                           

Note that the components of the magnetic field ${\bf B}^{(1)}$ and ${\bf B}^{(2\ast )}$ depend on the ``source" 
function $\Lambda _0$, which is determined by the obliquity, spin period (through $\theta _0$), magnetic colatitude 
and azimuth of a field-line foot point at the PC surface. For electrons accelerating from the PC along the 
favorably curved ($\cos \phi _{\rm pc}^0 \geq 0$) field lines, $\Lambda _0$ is always positive (we 
assume that $0\leq \chi < 90^{\circ }$, i.e. ``normal polarity" pulsar; see also MH04). However, for the unfavorably curved ($\cos \phi _{\rm pc}^0 < 0$) field lines 
there may be a situation where the second term in $\Lambda _0$ dominates. In this case (e.g. for the millisecond 
pulsars with high obliquities) the unfavorably curved field lines may become inefficient in providing continuous 
steady-state flow of electrons (see also Muslimov \& Harding 2003, MH04). 
This effect should be taken into account in calculating the 
3-D magnetic structure for the short-period pulsars. We should also point out that one of the advantages of having 
our analytic expressions for ${\bf B}^{(1)}$ and ${\bf B}^{(2\ast )}$ depend on the source function $\Lambda _0$ 
is that the latter can be replaced by $\varepsilon ~\Lambda _0$, where $0 < \varepsilon < 1$ is an arbitrary 
factor which takes into account the possibility that at the stellar surface the electron current may be a factor 
of $\sim ~ \varepsilon $ less than the Goldreich-Julian value. This is important for modeling the effect of 
relativistic electron flow of different magnitude on the magnetospheric structure. In other words, by using our 
model we can explore the space-charge-limited flow approximation (e.g. by measuring the magnitude 
of the effect of field line focusing and sweep-back in pulsars), examine the role of ${\bf E}\times {\bf B}$ 
drift in determining the magnetospheric structure at very high altitudes, and probe the occurrence of different 
acceleration conditions on favorably ($\cos \phi _{\rm pc}^0 \geq 0$) and unfavorably ($\cos \phi _{\rm pc}^0 < 0$)  curved field lines. We plan to address these and other consequences of our model, including possible observational tests, in subsequent studies.

\section{Discussion and Conclusions}

In this study we began a quantitative analysis of the distortion of open magnetic field lines of a rotating NS with an arbitrary obliquity angle in the presence of relativistic charge flow. The static field configuration is assumed to be dipolar. Our analysis is aimed at the derivation of simple analytic formulae which can be used up to very high altitudes, say $\sim 0.5-0.7$ of the light cylinder radius. We presented the explicit analytic expressions for the first and second order corrections to the static dipole magnetic field that are produced by the effect of relativistic flow of charged particles (e.g. with the negative net charge) along the open field lines, bulk magnetosphere rotation, and non-vanishing ${\bf E}\times {\bf B}$-drift of a net charge across the open field lines. For the longitudinal current, we employed the corresponding expression derived earlier in the space-charge-limited-flow approximation by MT92. The longitudinal current produces a substantial distortion of the open field lines at high altitudes by twisting them 
in a direction opposite to that of rotation (as viewed down the magnetic pole) and making the entire bundle of open field lines less flaring (more focusing) along the magnetic axis (see equations [\ref{B1r}]-[\ref{B1phi}]). The effect of field-line twisting is clearly seen for small obliquities (see e.g. Fig. 2, cases $\chi = 0^{\circ }$ and $30^{\circ }$), in which case it can also be recognized as the sweep-back effect. The effect of focusing of open field lines can be seen in Figure 1 (cases $\chi = 30^{\circ }$ and $60^{\circ }$). The magnitude of these effects is determined by the obliquity and spin-period, so that these effects are much more pronounced for small obliquities, in contrast to the vacuum case. The global current associated with the bulk magnetosphere rotation distorts the open field lines (see terms that are not proportional to $\Lambda _0$ in equations [\ref{B2r}]-[\ref{B2phi}]) in such a way as to bend (at very high altitudes) the bundle of open field lines down toward the rotational equator. Finally, the effect of ${\bf E}\times {\bf B}$ -drift of charges (of the same sign as the net charge of relativistic longitudinal flow) across the open field lines results in some asymmetry between the leading and trailing edges of the bundle of open field lines (see equations [\ref{B2*r}]-[\ref{B2*phi}]). The distortion of open field lines caused by this effect can be seen in Figure 1 (cases $\chi = 30^{\circ }$ and $60^{\circ }$).   

A number of models of emission from rotation-powered pulsars have relied on the magnetic field
structure of the vacuum retarded dipole (Deutsch solution).  The predicted high-energy pulse profiles in 
outer gap (e.g. Cheng et al. 2000), two-pole caustic (Dyks \& Rudak 2003, Dyks, Harding \& Rudak 2004) 
and slot gap (MH04) models, where most of the emission occurs in the outer magnetosphere, depend sensitively 
on the structure of the field at high altitude.  The large differences in the rotational distortions of the 
magnetic field in vacuum and non-vacuum cases, that is demonstrated by our solutions, could produce important 
changes in the predictions of such models.
 
Although the main area of application of our study was meant to be the modeling of particle acceleration and 
emission of energetic photons at high altitudes in the open-field line regions of pair-starved pulsars 
(especially in millisecond pulsars), the solution presented here should also be applicable to the majority of
pulsars producing high pair multiplicity in their magnetospheres.  This is because our solution assumes a
primary current given by a space-charge limited flow model.  Even though this model (and any acceleration model)
necessarily departs from the force-free (ideal MHD) case, the departure is small.  Furthermore, we have 
illustrated in previous studies (MH04, Muslimov \& Harding 2004b) that even in the charge-starved limit, the actual 
space-charge along the open field lines approaches the Goldreich-Julian charge, and thus the force-free condition, 
at high-altitudes in the magnetosphere.  Thus the difference between the high-altitude field structure of a
pair-starved pulsar and a pulsar producing an abundance of pairs should be minimal.

It is worth mentioning some other areas where our formulae can be used and subjected to observational 
tests. First of all, our analytic expressions can be implemented in the analysis of pulse polarization properties 
in radio pulsars to probe e.g. the geometry of the magnetic field lines in emission cites (e.g. Gangadhara \& Gupta 2001; 
Blaskiewicz, Cordes \& Wasserman 1991; Dyks, Rudak \& Harding 2004; and Hibschman \& Arons 2001). Second, the results of our study can be used for the interpretation of multi-frequency (e.g. in radio-, IR-, optical-, X-, and $\gamma $-rays) light curves of pulsars to explore the 3-D picture of pulsar emission (see Cheng et al. 2002; and Romani 2002 for a review). And third, our formulae might be useful in  modeling of magnetosphere geometry, in general, and particle acceleration regions, in particular, that is needed in interpretation of observations. 
For example, the recently discovered double pulsar system PSR J0737-3039 A\&B (Burgay et al. 2003; Lyne et al. 2004) may provide us with the opportunity to probe both the pulsar magnetospheric structure and wind density (see e.g. Arons et al. 2005; Rafikov \& Goldreich 2005; and Zhang \& Loeb 2004). Also, our study may be applicable to the interpretation of pulsar breaking indices (see e.g. Manchester \& Taylor 1977 for a general discussion). Finally, it is interesting to point out that in the course of pulsar evolution their spin periods and maybe field strengths (and/or inclination angles) as well as (as our present study implies) the geometries of their open-field line regions can change. This effect should be taken into account in population synthesis. 

In the future, we can extend our present study in, at least, three different ways: a) To derive the appropriate analytic solution valid beyond the light cylinder; b) To incorporate the deviation of open-field line structure 
at high altitudes into the electrodynamic model to derive a more accurate high-altitude $E_{\parallel}$; 
c) To proceed with detailed modeling of high-altitude pulsar emission and comparison with the observational data. 
Note that items b) and c) imply some additional 
development of our model, while item a) may turn into a separate study falling under the topic of a pulsar 
relativistic wind. In addition, after implementing item b) we will be able to perform well-founded validation of the existing scenarios of particle acceleration and emission in pulsars. Finally, as a result of effort c), we expect to come up with some constraints on any theoretical model of open field lines near the light cylinder.

\acknowledgments 
We acknowledge support from the NASA Astrophysics Theory Program through the Universities 
Space Research Association.

\clearpage 

\appendix
\section{Components of vectors $\bf g$ and $\bf h$}

In this Appendix we present explicit expressions for the spherical components of vectors $\bf g$ and $\bf h$, 
which are defined in Section 2.1 (see formulae [\ref{J-2}] and [\ref{curl-b2}]).

The $r-,~\theta -$ and $\phi -$ components of vector $\bf g$:  
\begin{eqnarray}
g_{\rm r} & = & {1\over \eta }~(\cos ^2 \chi~\sin ^2 \theta + \sin ^2 \chi~\cos ^2 \theta + 
\sin ^2 \chi~\sin ^2 \theta ~\sin ^2 \phi - \nonumber \\
& & 2~\sin \chi ~\cos \chi ~ \sin \theta ~\cos \theta~\cos \phi )~\cos \theta ,  
\label{gr}
\end{eqnarray}
\be
g_{\theta } = {1\over {2\eta }} (\cos \chi ~\sin \theta - \sin \chi~\cos \theta~\cos \phi)^2~\sin \theta,
\label{gtheta}
\ee
and
\be
g_{\phi } =  {1\over {2\eta }} (\cos \chi~\sin \theta - \sin \chi~\cos \theta~\cos \phi)~\sin \chi~\sin \theta ~
\sin \phi .
\label{gphi}
\ee
respectively.

The $r-,~\theta -$ and $\phi -$ components of vector $\bf h$:
\be
h_{\rm r}  = {2\over \eta }~\Lambda _0~\left[(1-\kappa ) \cos \chi~\cos \theta - {3\over 2}~
\theta _0~\xi ~{{\theta -\sin \theta~\cos \theta }\over {\sin \theta }}\sin \chi~\cos \phi \right] ,  
\label{hr}
\ee
\be
h_{\theta } = 0,
\label{htheta}
\ee
and
\be
h_{\phi } =  {3\over {\eta }}~\Lambda _0~\theta _0~\xi~{{2~\cos \theta +\theta ~\sin \theta -2}
\over {\sin \theta    }}~\sin \chi~\sin \phi .
\label{hphi}
\ee

\newpage

~
\vskip 4.0cm
\hskip -2.0cm
\includegraphics[width=200mm]{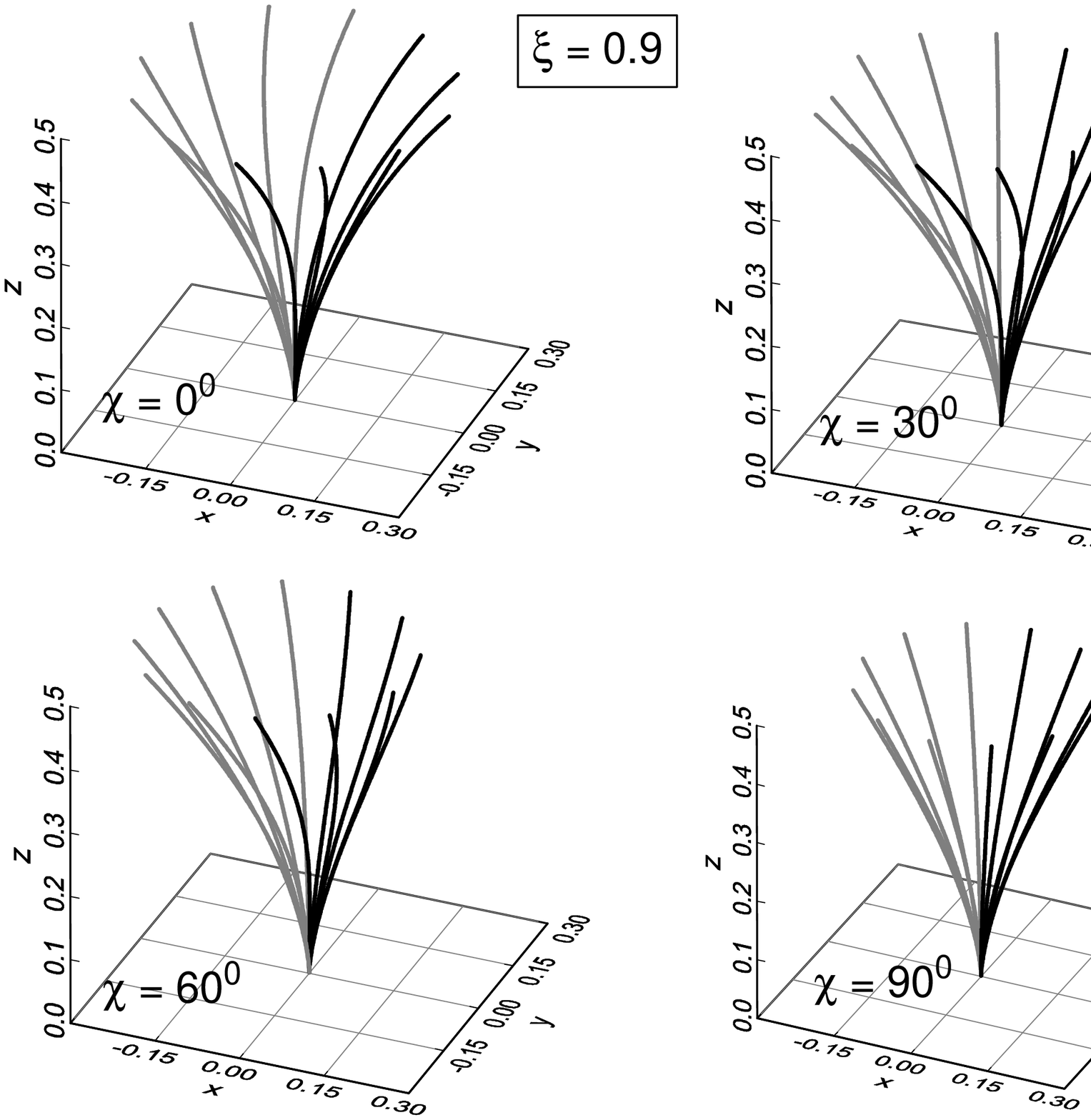}
\vskip -3.0cm
\figureout{f1.eps}{
The side views of a bundle of open field lines with $\xi = 0.9$, for different values of pulsar obliquity 
$\chi $ (= 0, 30, 60 and 90$^{\circ }$). The Cartesian coordinate system (X, Y, Z) is centered at the magnetic pole, with the Z-axis along the magnetic moment, and positive X-axis pointing to the rotation axis. The coordinate values are in units of the light cylinder radius. The calculations are performed for the pulsar spin period of 0.3 s.  
Black lines denote favorably curved field lines ($\cos \phi _{\rm pc}^0 > 0$) and gray lines denote unfavorably curved field lines ($\cos \phi _{\rm pc}^0 < 0$).
    }    

\newpage
\vskip 0.3cm
\hskip -2.5cm
\includegraphics[width=200mm]{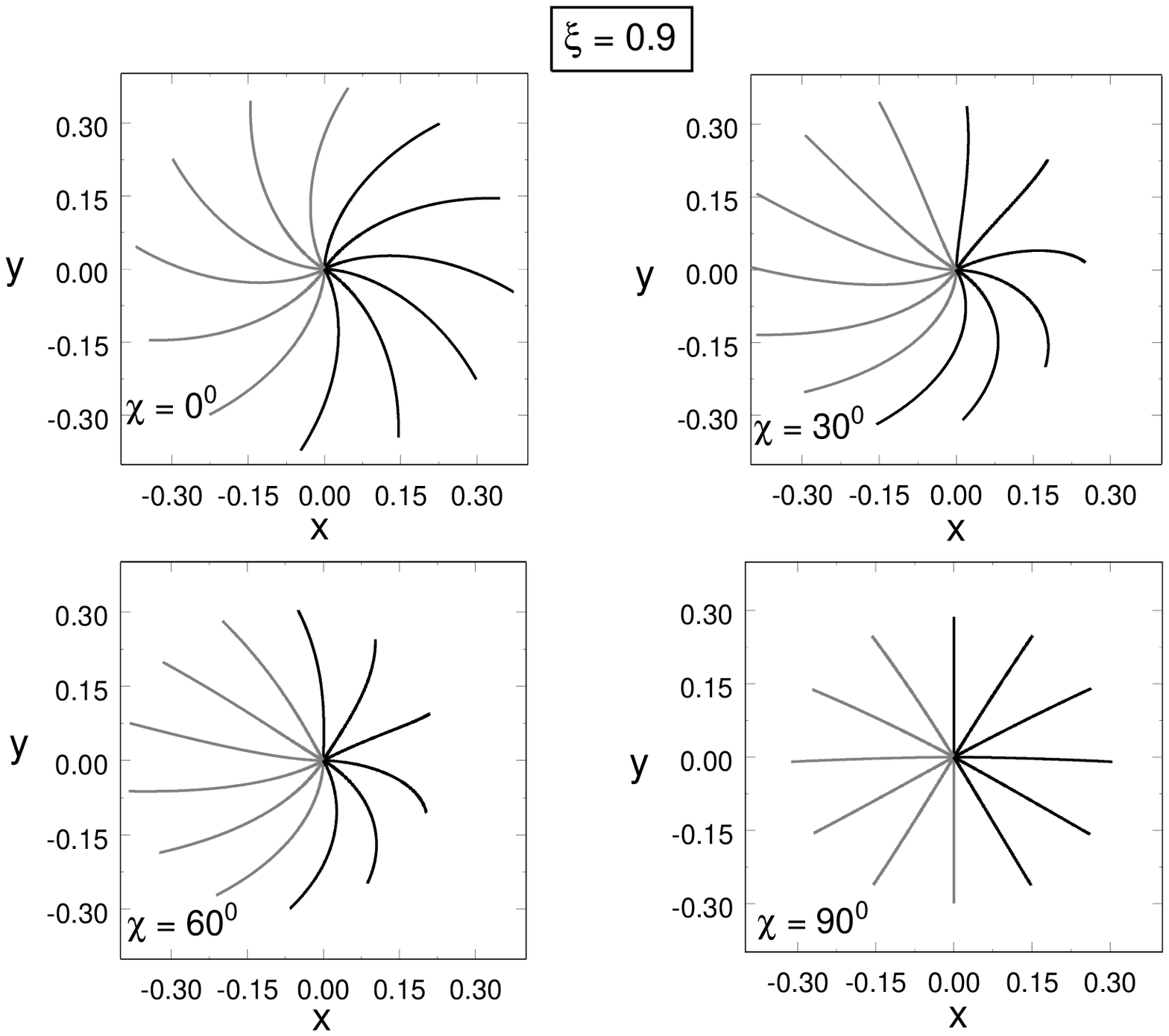}
\figureout{f2.eps}{
The views down the magnetic pole (from far above) of a bundle of open field lines with $\xi = 0.9$ for different 
values of pulsar obliquity $\chi $ (= 0, 30, 60 and 90$^{\circ }$). Same as in Figure 1. Leading field lines are at 
negative Y and trailing field lines are at positive Y.
    }    

\newpage
\hskip -2.0cm
\includegraphics[width=200mm]{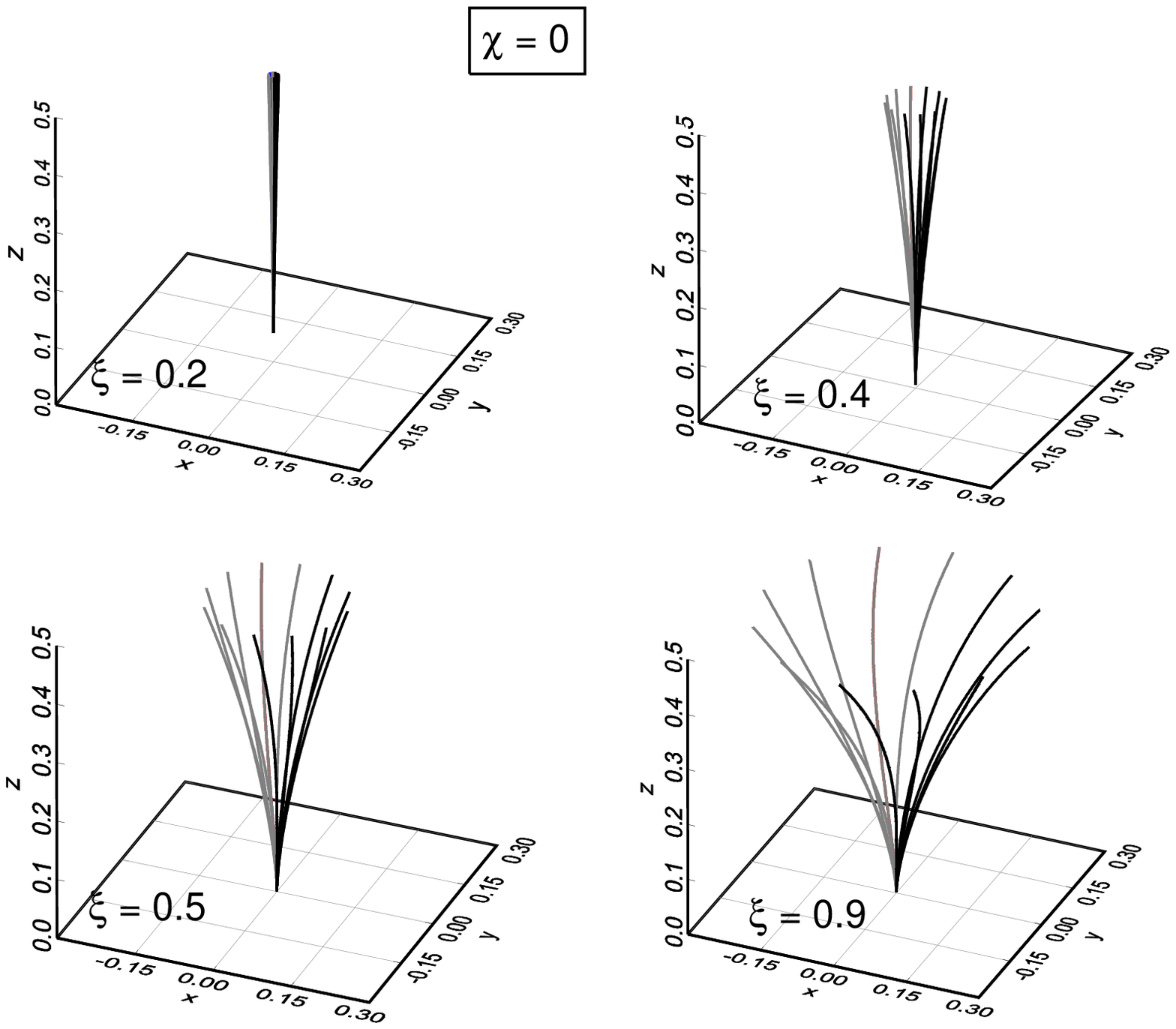}
\figureout{f3.eps}{
The side views of bundles of open field lines with $\xi$ = 0.2, 0.4, 0.5, and 0.9 for the aligned case ($\chi = 0$). Same as in Figure 1.
    }    

\end{document}